\documentclass[12pt]{iopart}
\usepackage{iopams}  
\usepackage{amssymb,graphicx,caption}
\usepackage{xfrac}
\usepackage{lineno}
\usepackage{color}
\usepackage{multirow}
\usepackage{epstopdf}
\usepackage{cite}
\usepackage{setspace}
\begin{document}
\title[Gluing Two Ag Nanoparticles]{Static and Ultrafast Optical Response of Two Ag Nanospheres Glued by a CdTe Quantum Dot.}

\author{Sabina Gurung $^{1,2}$ ,Asha Singh$^{1}$, Durga Prasad Khatua$^{1,2}$, Himanshu Srivastava$^{3}$, and J. Jayabalan$^{1,2,*}$}
\address{$^1$Nano Science Laboratory, Materials Science Section, Raja Ramanna 
	Centre for Advanced Technology, Indore, India - 452013.}
	\address{$^2$ Homi Bhabha National Institute, Training School Complex, 
	Anushakti Nagar, Mumbai, India - 400094.}
       \address{$^3$Lithography and Microscopy Laboratory, Synchrotrons Utilization Section, Raja Ramanna 
	Centre for Advanced Technology, Indore, India - 452013.}
	
\ead{*jjaya@rrcat.gov.in}


\begin{abstract}
A hot-spot of high local field can be created in between two closely placed metal nanoparticles 
by irradiating them with an appropriate wavelength and polarization of incident light. The strength of the field 
at the hot-spot is expected to get enhanced by even up to six orders of magnitude than 
that of the applied field. Placing a semiconductor quantum dot or an analyte molecule at the hot-spot 
is an essential step towards harnessing the enhanced field for applications. In this article, 
we show that it is possible to position a CdTe quantum dot (QD) between two larger silver 
nanospheres in colloidal solution. The extinction spectra measured during growth suggests that 
the final hybrid nanostructure have two touching Ag nanoparticles (NPs) and a CdTe QD in 
between them close to the point of contact. Using ultrafast transient measurements, it has been 
shown that the presence of CdTe QD strongly influence the dynamics when the probe  
excites the hot-spot. The method demonstrated here to place the semiconductor 
QD in between the two Ag NPs is an important step in the area of colloidal self-assembly and 
for application of hot-spot in plasmonic sensing, optoelectronics, energy-harvesting, 
nanolithography, and optical nano-antennas.
\end{abstract}

\section{Introduction}
When excited at localized surface plasmon resonance (LSPR) the metal nanoparticle (NP) can 
significantly enhance the strength of electromagnetic field around 
them\cite{Absorption_Scattering_small_particles_borhen_1983, Nanoplasmonics_enhancement_Nanotechnology_2006}. Such 
field enhancement can be as large as 20 times than that of the applied\cite{LSPR_AgNP_JPCB_2003}. 
This property of metal NP opened up its applications in plasmonic sensing, 
optoelectronics, energy harvesting, nanolithography, nano-antennas and 
imaging\cite{Plasmonics_solar_cell_optic_express_2008, LSPR_Bisosensing_Ruemmele_ACS_2010, 
	PLasmonic_application_nature_review_Harry_2010, Plasmon_solar_energy_Scott_2013, 
	metal_sc_plasmon_enhanced_Jiang_Adv_matter_2014, Plasmonic_photovoltaics_Fan_Molcules_2016, Non_symmetric_hybrid_Yinghui_Naturalscience_2017, LSPR_double_metal_complex_Heesang_Sensors_2018, Xu2013}.
A second generation of local field enhancement is realized when two metal NPs are placed 
close to each other with separation less than few nanometers. The field in between these particles 
can be enhanced by a factor of up to six orders of magnitude than that of the applied. 
Theoretical studies on generation of such high field regime, known as the hot-spot, using two 
metal nanoparticles of various size, shape and orientation is being studied by various
groups\cite{EM_Around_Ag_dimers_JCP_2004, LSPR_review_Analytica_chimica_2011}. 
In past, bottom up techniques such as chemical synthesis was used for the preparation of large 
quantity of isolated metal nanoparticles of different shapes with good size and crystal
quality\cite{Tailoring_Properties_crystanality_Nature_2007, SERS_Agnanocubes_CPL_2010,Luo2019}. To generate 
hot-spots, these particles were simply aggregated in the colloid. Such random hot-spots in
aggregates were also used for increasing the SERS signals from 
molecules\cite{SERS_Agnanocubes_CPL_2010}. Recently dimer shaped 
Ag nanoparticles, which can show hot-spot, was prepared in colloidal form by controlled addition 
of linking polymer or salt to Ag nanosphere colloid. When mixed, the linker molecule acts as glue 
to stick the two larger sized metal NPs. Because this glue is tiny, further attachment of metal NP 
to the same spot is prevented, creating the desired dimer structure\cite{Feng2012metalmetal, Mark2013dimer}. 
There are other similar solution based techniques to prepare dimers\cite{Dimer_field_enhancement_SERS_Nanoletter_2009,Fan2010Science}.
All these dimer preparation methods require high level of control over the number of linking molecules verses the number of nanoparticles to avoid formation of large aggregates. Although such techniques allow preparation of hot-spot, placing a semiconductor QD or an analyte molecule in that location is also important for harnessing the enhanced local field
for applications. Once placed at the hot-spot, the material will experience a strong field leading 
to enhancement in its absorption, emission and pronounced Fano-like effect and so forth\cite{Absorption_enhancement_NanoLett_2012, Fano_dimers_Manjavacas_NanoLett_2011}. 

In this article, we show that by mixing Ag NP colloid with TGA capped CdTe QDs in a particular
mixing ratio, it is possible to prepare hybrid nanostructure with CdTe QD stuck in between two Ag NPs. 
By comparing the measured time evolution of extinction spectra during self-organized growth
of the hybrid nanostructure with that of the calculated, we explain the process of hybrid
formation. Ultrafast pump-probe measurements can provide 
quantitative information regarding complex processes that take place between the metal and 
semiconductor nanocomposites. Using ultrafast transient transmission measurements it has been 
shown that the carrier dynamics in the hybrid colloid is effected only when probed by a 
light that generates high field in between Ag NPs. The method demonstrated here to place the 
semiconductor QD in between the two Ag NPs is an important step in the area of colloidal 
self-assembly and for application of hot-spot in plasmonic sensing, optoelectronics, energy -harvesting, nanolithography and optical nano-antennas.

\section{Experimental Details}

\begin{figure}[ht] 
	\centerline{\includegraphics[width=0.95\columnwidth]{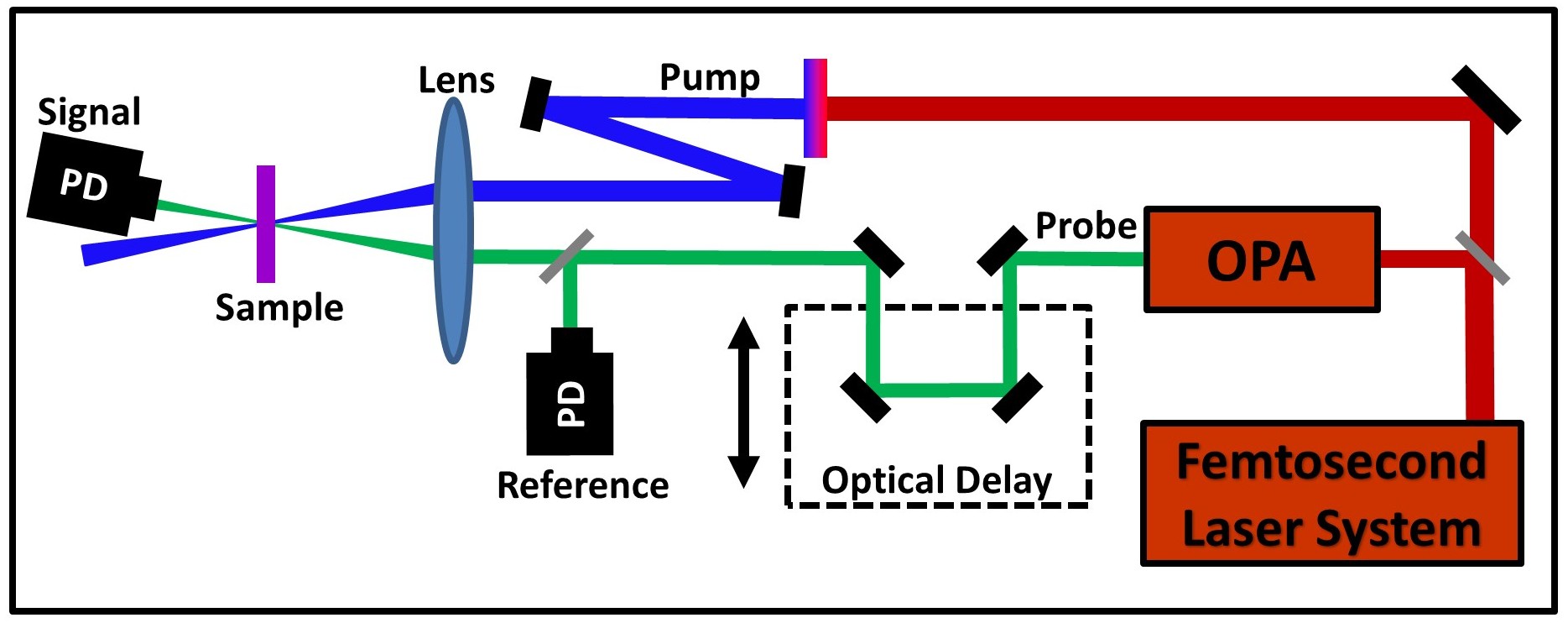}}
	\caption{Schematic of the pump-probe setup used for the measurement of transient transmission from the samples. \label{Fig:Setup}}
\end{figure}

The individual constituents of the hybrid nanostructure, citrate capped silver NPs and TGA capped CdTe 
QDs dispersed in water, were prepared separately by wet chemical 
methods\cite{Silver_sphere_Science_2001, Guo-ReactionConditions-TGA-CdTeNP-JPCB-2005, 
Abhijit-CdTe-Pre-PL-JPCC-2008}. The procedure followed for the preparation of this individual colloids 
has been reported earlier\cite{Sabina-JAP-2018}. Ag-CdTe hybrid colloid, labeled H$_{\gamma}$, is 
prepared by mixing $V_p$, volume of Ag NP colloid and $V_d$, volume of CdTe QD colloid. Here the mixing 
ratio, $\gamma$, is defined as $V_p/V_d$. In the mixed solution the Ag-CdTe hybrid nanostructures are 
expected to form because of self-organized growth\cite{Sabina-JAP-2018, 
metal_Sc_hybrid_selforganization_exciton_plsmon_Strelow_2016, 
Ag_cdte_selforganized_electrostatic_interaction_Wang_Spect_Acta_2005, Luo2019}.
A 1 kHz, 35 femtosecond oscillator-amplifier system was used for the transient 
transmission measurements in standard pump-probe geometry (Fig.\ref{Fig:Setup}). The pump beam (400 nm) 
for the experiment was obtained by generating second harmonic of a part of the 800 nm beam from the 
laser system using a BBO crystal. The 800 nm beam was also used for pumping an optical parametric 
amplifier (OPA). A fraction of the OPA output beam (408 nm or 550 nm) was used for probing the optical 
response of the sample. The polarization of the 408 nm and 550 nm beams 
were parallel and perpendicular to that of pump beam respectively. The sample was constantly circulated 
in a 1 mm cell to avoid any thermal damage.

\section{Results and Discussion}

\begin{figure*}[ht] 
	\centerline{\includegraphics[width=0.95\textwidth] {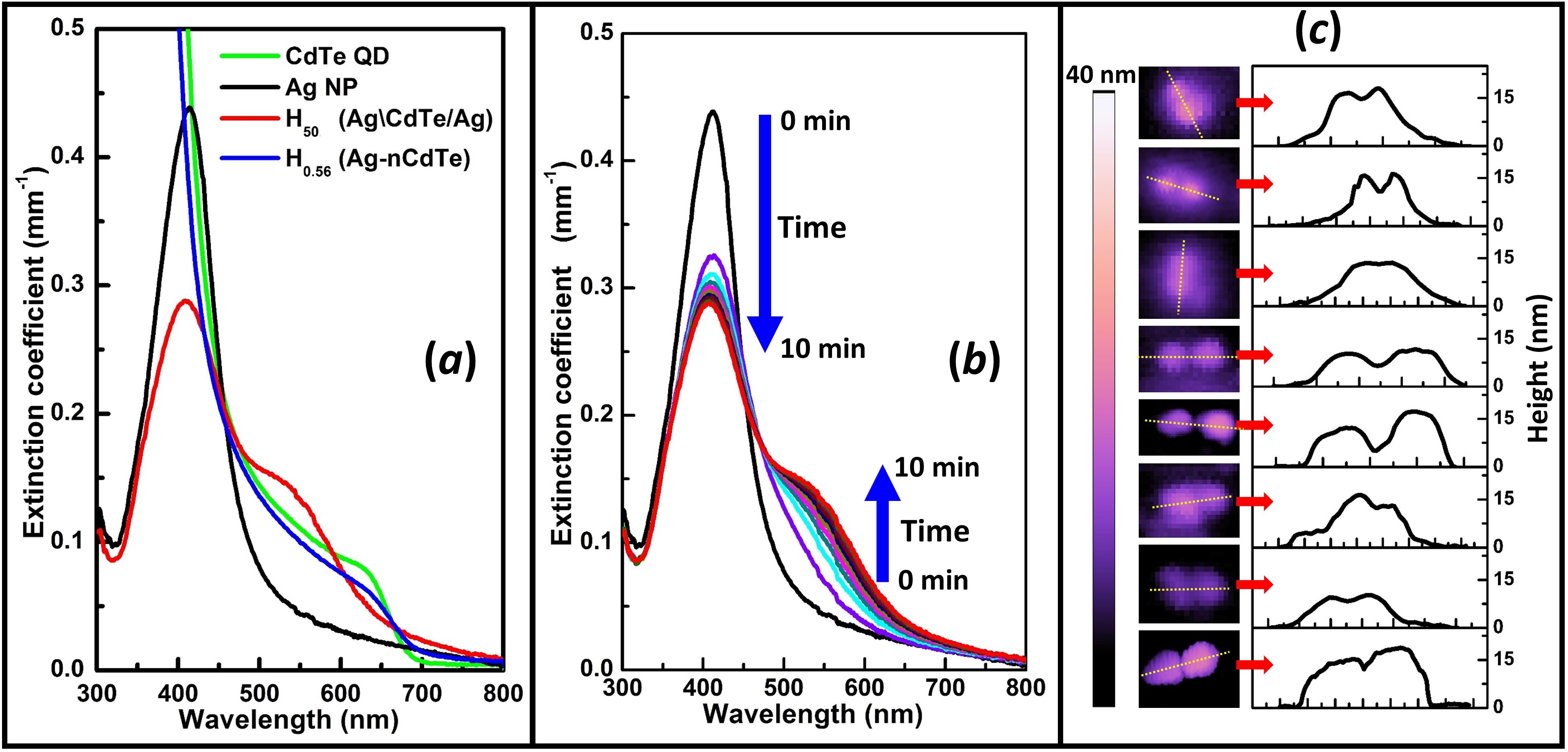}}
	\caption{Static optical and structural characterization of individual, mixed and hybrid colloids:
		({\it a}) Extinction coefficient of CdTe QD, Ag NP, H$_{50}$ hybrid and H$_{0.56}$ hybrid colloids. 
		({\it b}) The time evolution of extinction spectrum of the mixed colloid from just after mixing to
		about 10 min. ({\it c}) Representative AFM images of hybrid nanostructures present in the H$_{50}$
		hybrid colloid. $x$ is the distance along the substrate surface.
		\label{Fig:Characterization}} 
\end{figure*}

Figure.\ref{Fig:Characterization}{\it a} shows the extinction spectrum of the Ag NP colloid 
and CdTe QD colloid. The extinction spectrum of the Ag NP colloid shows a single peak at 413 nm which 
corresponds to the LSPR of small silver nanospheres dispersed in 
water\cite{Samanta-ultrafast-AgCdTe-JPCC-2016, conc_Ag_UV_VIS_Paramella_2014}. The extinction spectrum of the 
CdTe QD colloid shows exciton peak at 635 nm which is due to 1s(e)-1s(h) 
transition\cite{Luminescent_Cdte_QD_2003, Pengs_cdte_quantum_dots_size_2003, UCPL_CdSe_2005}. For structural 
characterization few drops of these samples were dried on a mica substrate and the topography of 
the surface was measured using Atomic Force Microscope (AFM). Some representative AFM images of the 
individual colloids are shown in Fig.S1 of supporting information (SI). The image that 
corresponds to individual Ag NP colloid and CdTe QD colloid shows presence of mostly well 
separated particles. The average diameter of the Ag nanoparticles and CdTe QDs measured using the 
height profile from these AFM images are 17 nm and 3.5 nm respectively.

\begin{figure}[ht] 
	\centerline{\includegraphics[width=0.95\columnwidth] {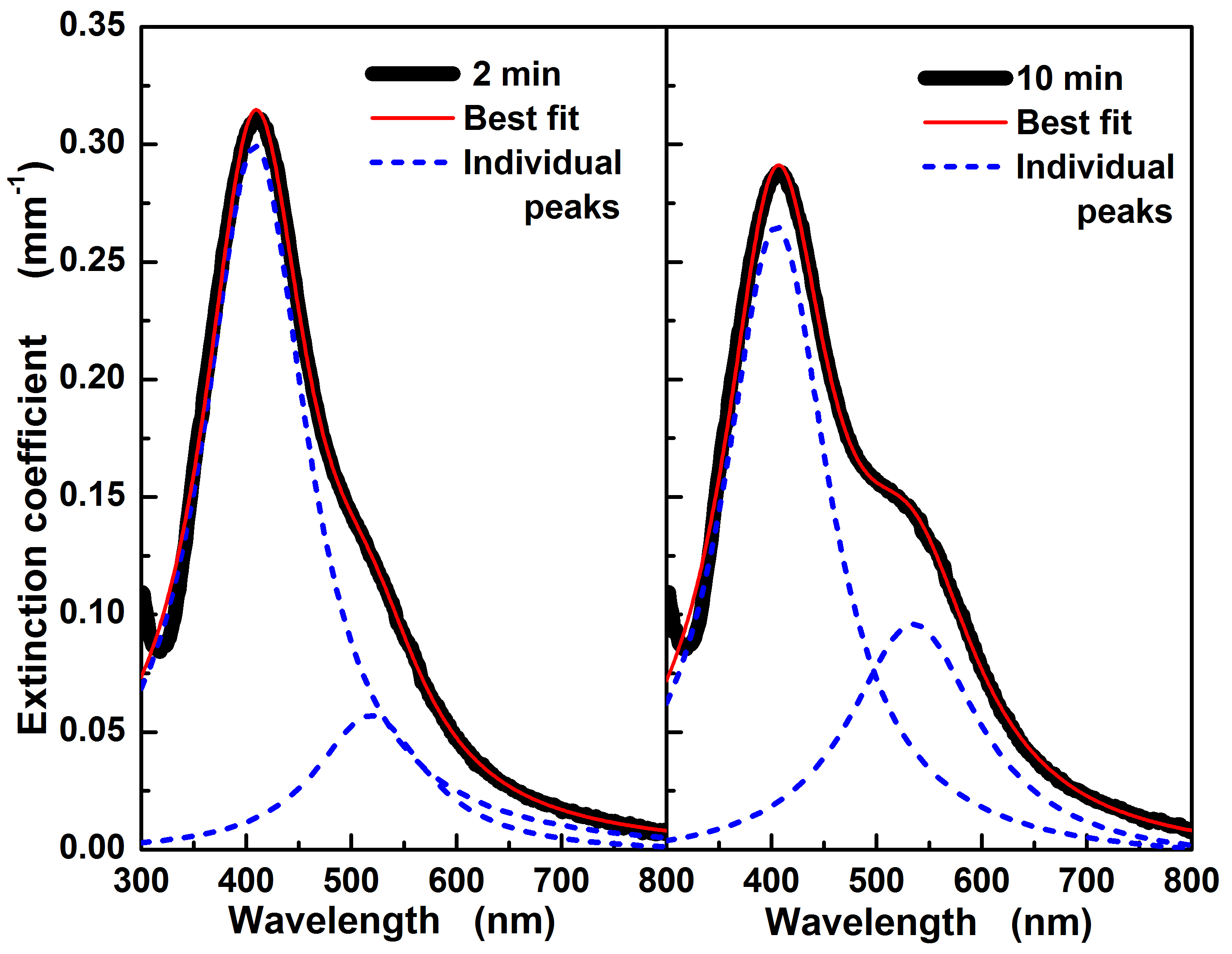}}
	\caption{Extinction spectrum of the mixed sample (black lines) after ({\it a}) 2 min and ({\it b}) 
		10 min. Best fit to these experimental spectra using two Gaussian peaks (red lines). The 
		contribution from individual peaks are also shown (blue lines). \label{Fig:PeakShift}}
\end{figure}

Ag-CdTe hybrid sample, H$_{50}$ was prepared by mixing CdTe QD colloid with Ag NP 
colloid at a mixing ratio of $\gamma$ = 50. Immediately after the addition, the LSPR peak 
red-shifts slightly and at the same time the extinction peak height reduces to about 74\% of that before addition 
(Fig.\ref{Fig:Characterization}{\it b}). Although the red-shift is very small, repeated preparation  
shows that the red-shift do occur at about 1 min after mixing (Fig.S2). Over the period of next few 
minutes another peak appears on the red-side. For example, the spectrum of the sample 
taken after 2 min could be fit well with two Gaussian peaks (Fig.\ref{Fig:PeakShift}{\it a}). 
From 2 min to 10 min, the original LSPR peak blue-shifts and continues to reduce in strength. Whereas
the longer wavelength peak red-shifts and grows (Fig.\ref{Fig:Characterization}{\it b}). 
After about 10 minutes, the mixed colloid shows no further change in the spectrum. The final stable 
hybrid sample thus formed (H$_{50}$) has peaks at 410 nm and 535 nm with strength 65\% 
and 32\% compared to that of bare Ag NP colloid respectively. Note that, the sample would be continuously 
evolving during the measurement of the spectra. Nevertheless, the measured spectra can be used for gaining 
insight into the hybrid formation process.

A self-organized growth of Ag-CdTe hybrid nanostructure is expected in the mixed sample because of the tendency of 
TGA to get attached on the surface of Ag NP\cite{Ligand_exciton_photovoltaic_PbS_Jin_PCCP_2017, 
TGA_CdTe_Europium_surface_coordinated_emission_Gallagher_inorganic_chemistry_2013, Sabina-JAP-2018}. 
As the TGA is already attached to the CdTe QD, these QDs are also brought in close proximity to the Ag NP\cite{Ligand_exciton_photovoltaic_PbS_Jin_PCCP_2017, 
	TGA_CdTe_Europium_surface_coordinated_emission_Gallagher_inorganic_chemistry_2013, Sabina-JAP-2018}. The final 
structure of hybrid formed by this self-organized growth process depends on the ratio between number of 
Ag NPs and CdTe QDs in the mixture, properties of the capping agents, size and shape of the individual 
particles. When Ag NP and CdTe QD colloids are mixed in a ratio of $\gamma$ = 0.56, hybrid nanostructures 
of Ag NP surrounded by nearly 45 CdTe QDs (H$_{0.56}$ or Ag-nCdTe) are formed\cite{Sabina-JAP-2018}. Once 
the CdTe QDs surrounds the Ag NP completely, it prevents further attachment of CdTe QD on to the Ag NP thus 
making the final hybrid structure stable. The extinction spectrum of the H$_{0.56}$ sample thus formed is also 
shown in Fig.\ref{Fig:Characterization}{\it a} for comparison. Clearly the extinction spectrum of this
H$_{0.56}$ sample is very different from that of the H$_{50}$ sample. The H$_{0.56}$ has an increasing  
absorption strength towards UV similar to that of bare CdTe QD colloid. On the other hand the 
H$_{50}$ sample shows a dip in absorption around 320 nm which resembles more of Ag NP colloid. The AFM 
images of the H$_{50}$ sample dried on mica shows several twin particles of each having nearly 17 nm 
height (Fig.\ref{Fig:Characterization}{\it c}). Based on the height, each of these particles should be the Ag NPs 
present in the mixed colloid.

The mixing ratio used for the sample H$_{50}$ shows that there are 
nearly two Ag NPs for each CdTe QD. The number density of Ag NPs and CdTe QDs in the individual colloids 
were also estimated by using the numerically calculated extinction cross-section of a single Ag NP and CdTe
QD and comparing it with the corresponding experimental extinction spectrum. The number densities estimated 
by using this procedure is also in agreement with the 2:1 number ratio of Ag NPs and CdTe QDs. Based on the 
number density estimation and the tendency of TGA to get attached to the Ag 
surface\cite{Ligand_exciton_photovoltaic_PbS_Jin_PCCP_2017, 
	TGA_CdTe_Europium_surface_coordinated_emission_Gallagher_inorganic_chemistry_2013}, we propose that in 
the final hybrid colloid, there are nanostructures with two Ag NPs attached to a single 
CdTe QD. 

If a mixed hybrid sample was prepared with mixing ratio $\gamma$ higher than 50, stable colloid do 
get formed. The final extinction spectrum of these hybrid colloids ($\gamma > 50$) has an extinction 
spectrum that looks similar to that observed in the intermediate stages (Fig.\ref{Fig:Characterization}{\it 
b}). For these $\gamma$ values, there will not be enough CdTe QDs to join all the Ag NPs leaving several 
unattached Ag NPs along with few Ag\textbackslash CdTe/Ag hybrids\cite{Sabina-JAP-2018}. On the other 
hand, if the mixing ratio is lower than 50, aggregated nanostructures were found to settle 
down at the bottom of the sample. This can be explained by the presence of excess CdTe QDs which can 
link several Ag\textbackslash CdTe/Ag NPs forming large chain-like structures which eventually becomes 
heavy enough to settle down\cite{Aggregated-JPCM-2008, Removal_Ag-NP-Advances-in-NP-2019}. 

\begin{figure} 
	\centerline{\includegraphics[width=0.95\columnwidth]{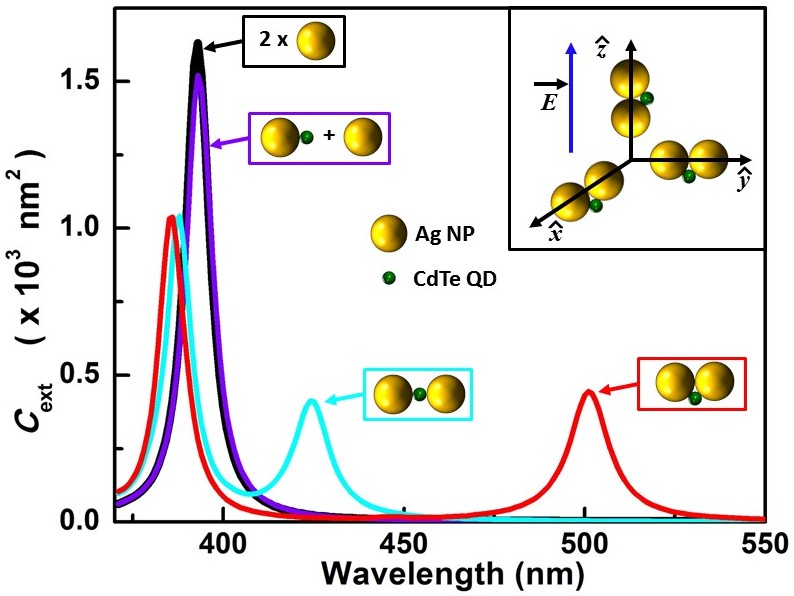}}
	\caption{Calculated extinction cross-section of two non-interacting Ag NPs, attached Ag-CdTe 
particles, chain of Ag-CdTe-Ag particles and touching Ag-Ag particles with CdTe using T-matrix. 
\label{Fig:CalculatedSpectra}}
\end{figure}

To understand the time evolution of the measured extinction spectra after mixing and to further 
understand the structure of the final hybrid formed, optical response of various Ag-CdTe hybrid 
nanostructures were numerically calculated using T-matrix technique (Fig.\ref{Fig:CalculatedSpectra}). 
T-matrix is a numerical method for computing optical response of a collection of spherical 
particles\cite{Light-scattering-randomly-symmetric-optical-society-1991, 
T-matrix-ensembleof-spheres-optical-society-1996}. For Ag NPs of size of the order of 17 nm quantum confinement
effects can be neglected and hence experimental bulk dielectric constants were used for the optical response 
calculations\cite{Optical-constant-JohnsonChristy-prb-1972}. On the other hand, dielectric constant of a CdTe QD 
of size 3.5 nm will be substantially different from that of its bulk because of the quantum confinement 
effect\cite{Dielectric-function-CdTe-trial-error-JPCC-2010}. The dielectric constant of CdTe QD was estimated 
using the method reported by Marcelo Alves-Santos {\it et al.} which uses the measured extinction 
spectrum and a trial and error procedure\cite{Dielectric-function-CdTe-trial-error-JPCC-2010}. In the 
colloidal solution, it is expected that particles will be oriented in all possible directions. To mimic 
that situation with minimal computational cost, the extinction spectrum is estimated by averaging the 
responses calculated for a given field direction and particles aligned along the three Cartesian directions 
(see inset of Fig.\ref{Fig:CalculatedSpectra})\cite{Optical-proeprties-metal-Kelly-JPCB-2003, 
	T-matrix-ensembleof-spheres-optical-society-1996}.

The extinction cross-section of the Ag NP shows a single peak at 393 nm (Fig.\ref{Fig:CalculatedSpectra}). 
In order to compare with the other structures where two Ag NPs are involved, $C_{ext}$ of the two isolated
Ag NPs (same as cross-section of single Ag NP multiplied by two) is plotted in the 
Fig.\ref{Fig:CalculatedSpectra}. Once the Ag NP and the CdTe QD 
colloids were mixed, the most possible situation would be that a single CdTe QD will get attached to a 
single Ag NP. The second Ag NP would not have joined yet. To mimic such situation we have calculated 
$C_{ext}$ of a single CdTe QD attached with an Ag NP and another Ag NP placed sufficiently away from it. 
The $C_{ext}$ thus obtained shows a red-shift of about 1 nm when compared to that of isolated Ag NP. 
Further, the peak height also reduces to about 90\% of that of two isolated Ag NPs. Attaching a higher 
refractive index material to a metal nanoparticle is known to red-shift its LSPR 
peak\cite{Dielectric-function-plasmon-ZPB-1975, Local-refractive-plasmon-Nanoletter-2003}. The changes in 
$C_{ext}$, the red-shift and the reduction in peak strength, when CdTe is attached to one of the Ag NP 
matches well with the observed change in the extinction spectrum measured nearly a minute after mixing
Ag NP and CdTe QD colloids (Fig.\ref{Fig:Characterization}({\it b})). When another Ag NP is attached to 
the other end of CdTe QD which is already attached to an Ag NP, it will result in the formation of a linear 
Ag-CdTe-Ag structure. The $C_{ext}$ of this structure shows a new peak at 424 nm and the original LSPR 
peak which was at 393 nm blue-shifts by 5 nm to 388 nm. These LSPR peaks, 424 nm and 388 nm appear 
when, field is aligned along the line joining the centers of the two Ag spheres (longitudinal axis) 
and perpendicular to it (transverse axis) respectively (Fig.\ref{Fig:CalculatedSpectra}). Fitting the 
measured extinction spectrum at 2 min after mixing shows a double peak structure which matches well 
with that of $C_{ext}$ of Ag-CdTe-Ag (Fig.\ref{Fig:PeakShift}). The final spectrum of 
the hybrid has a second peak at 535 nm which is much more red-shifted compared to that calculated for the 
Ag-CdTe-Ag structure. Although capping prevents Ag NPs to join together, once brought in close proximity 
to each other, the metal particles do tend to join and restructure themselves\cite{Kinetics-Ag-dimer-JACM-2011, 
Built-in-mechanism-room-temperature-ACS-Nano-2011, Merging_nanoparticles-wettability-AgNP-Nature_communication-2013}. 
Thus, once joined by the CdTe QD it is possible that over time the linear Ag-CdTe-Ag hybrid changes its 
structure to that with two Ag NPs touching each other with CdTe QD pushed to one side (Ag\textbackslash CdTe/Ag) 
as shown in Fig.\ref{Fig:CalculatedSpectra}. The calculated $C_{ext}$ of Ag\textbackslash CdTe/Ag hybrid 
match closer to the experimentally observed final extinction spectrum which has peaks at 408 nm and
535 nm (Fig.\ref{Fig:Characterization}{\it a}). Thus the peak at 408 nm corresponds to the LSPR originating
when the field is aligned along the transverse axis (T-LSPR). Similarly the 535 nm peak corresponds to the 
LSPR when field is aligned along the longitudinal axis (L-LSPR).

\begin{figure} 
	\centerline{\includegraphics[width=0.95\columnwidth]{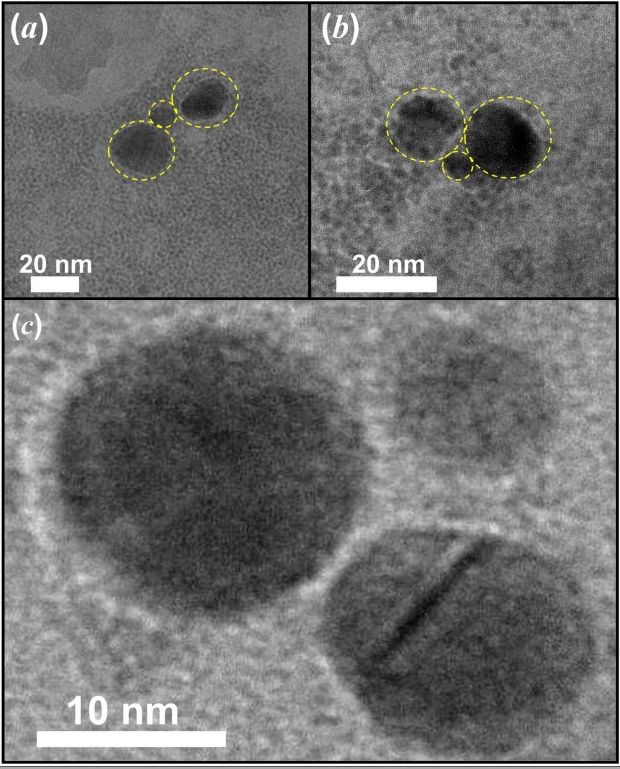}}
	\caption{TEM images of Ag\textbackslash Cdte/Ag hybrid nanostructures. 
		\label{Fig:TEM}}
\end{figure}

Comparing the measured temporal evolution of the extinction spectra (Fig.\ref{Fig:Characterization}{\it b})
and the calculated optical responses of the Ag NP and hybrid nanostructures (Fig.\ref{Fig:CalculatedSpectra}), 
the growth of the final hybrid nanostructure can now be explained. Just after mixing, CdTe QD 
gets attached to Ag NP forming a large number of hybrid nanostructures each with one Ag NP and one 
CdTe QD. Such Ag-CdTe hybrid structure coexist in the colloidal solution along with almost equal number 
of unattached Ag NPs. This is indicated by the observed small red-shift and reduction in strength of 
the LSPR peak when measured 1 min after mixing (Fig.S2). As time progresses, another 
Ag NP also gets attached to the Ag-CdTe nanostructure to form an Ag-CdTe-Ag linear hybrid nanostructure. 
The observed second peak on the red-side of the extinction spectra measured in the first few minutes 
indicates the formation of such structures. Followed by this, the linear Ag-CdTe-Ag structure realign 
themselves to form Ag\textbackslash CdTe/Ag structure. This explains the observed red-shifting of the 
long wavelength peak which finally settles down at 535 nm. To further confirm these observations we have also
prepared Ag\textbackslash CdTe/Ag hybrids with larger Ag NPs with diameter nearly 50 nm (Fig.S4). Once 
again the experimental results matches well with that of the calculated using T-matrix. 
Thus the final stable colloid, H$_{50}$ should have hybrid nanoparticles of Ag\textbackslash CdTe/Ag 
structure. The size of the CdTe QD is much smaller and is now in between two large Ag NPs, 
further attachment of Ag NP to the same CdTe QD is prevented. 
Because of the limitation in resolution, the AFM image could not resolve the presence of small CdTe QD
sitting in between the larger Ag NPs. Thus the AFM images of the Ag\textbackslash CdTe/Ag 
hybrid show only a Ag NP dimer-like structure (Fig.\ref{Fig:Characterization}{\it c}). 
In some of the AFM images of H$_{50}$, we find few hybrid structures with Ag NPs attached 
to each other in triangular formation and few Ag NPs in chain-like structure (inset of Fig.S3). 
The T-matrix calculation shows that these triangular and chain-like structures have a very 
different extinction spectra compared to that observed in the experiment (Fig.S3). 
Transmission electron microscope (TEM) could also be used for studying the structure of final
self-aggregated Ag\textbackslash Cdte/Ag nanostructures. Figure.\ref{Fig:TEM} shows some of the
TEM images of the colloid containing Ag\textbackslash Cdte/Ag hybrid nanostructures. The 
presence of the capping agents (polymers) in hybrid sample makes the process of capturing 
TEM images very difficult. Removal of these polymer by centrifuging would also destroy the 
hybrid nanostructure which is present in the colloidal solution. Further, during TEM studies, 
the measurement had to be performed at very low irradiation and the images had to be accrued 
quickly. Sometimes during the measurement, we still find that the particle do disappear during 
image accusation. The TEM images of Ag\textbackslash Cdte/Ag hybrid nanostructures clearly
shows presence of a smaller particle (size in the range of CdTe QDs) in between two larger 
sized particles (size of the order of Ag NPs) which matches well with that of expected from 
optical characterization (Fig.\ref{Fig:TEM}).

It is interesting to note that the observed strengths of the LSPR peaks in te case of Ag\textbackslash CdTe/Ag 
hybrid colloid can be explained by a simple model based on the
orientation of the particles. Because of spherical symmetry, all the Ag NPs in the colloid absorbs the incident 
light irrespective of its polarization. Although in a colloidal sample all orientations of 
Ag\textbackslash CdTe/Ag nanoparticle is possible, consider a simpler picture where all the particles are 
randomly aligned parallel to one of the three Cartesian directions with equal probability (inset of 
Fig.\ref{Fig:CalculatedSpectra}). Such assumption in the distribution of particles has been shown to match
well with that of the measured optical response of colloid containing randomly oriented ellipsoids and 
nanoprisms\cite{Optical-proeprties-metal-Kelly-JPCB-2003, Aggregated-JPCM-2008}. Let the electric field be aligned 
to one of the Cartesian direction, about 66.7\% ($\sfrac{2}{3}^{rd}$) of the particles will contribute 
to the T-LSPR peak (at 410 nm) because for these particles the polarization of light is oriented along the 
transverse axis. Whereas the rest 33.3\% ($\sfrac{1}{3}^{rd}$) of the particles has their longitudinal
axis aligned along the polarization and will show resonance at L-LSPR (535 nm). The measured LSPR peak 
heights for Ag\textbackslash CdTe/Ag hybrid colloid at 410 nm and 535 nm are 65\% and 32\% of that of Ag NP colloid 
at its LSPR peak which are very close to that predicted by the model. This also implies that the $C_{ext}$
of a Ag\textbackslash CdTe/Ag particle is exactly same as that of two isolated Ag NPs if polarization is
appropriately aligned. The reduction in $C_{ext}$ of the Ag\textbackslash CdTe/Ag hybrid colloid is mainly 
caused by the random orientations of the particles.

\begin{figure}[h] 
	\centerline{\includegraphics[width=0.95\columnwidth]{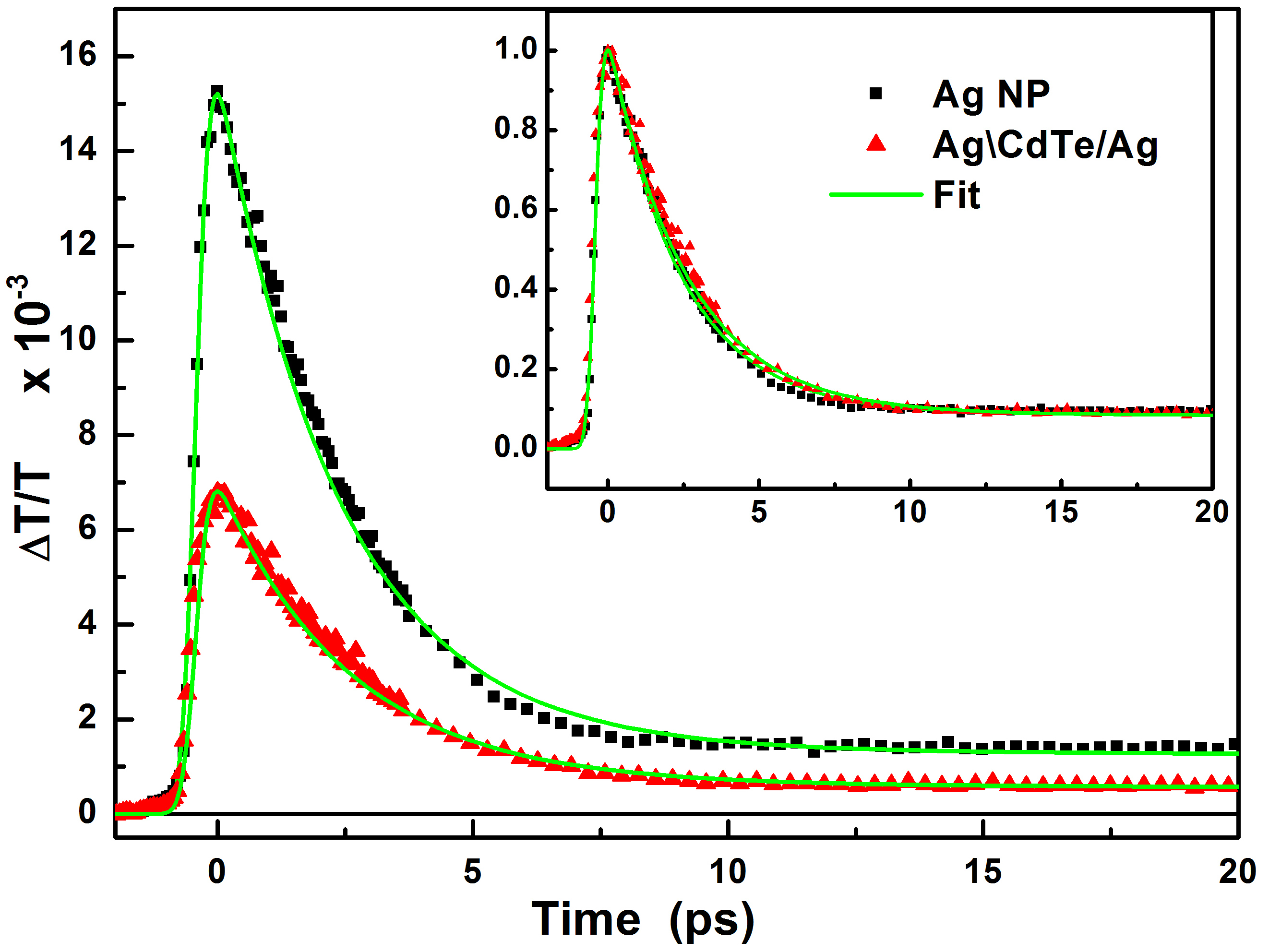}}
	\caption{Transient  curve of Ag NPs and Ag\textbackslash CdTe/Ag hybrid  colloidal samples excited 
at 400 nm and probe at 408 nm. Inset shows the normalized transient curve of these samples.	
\label{Fig:Transient410nm}}
\end{figure}

\begin{figure}[h] 
 	\centerline{\includegraphics[width=0.95\columnwidth]{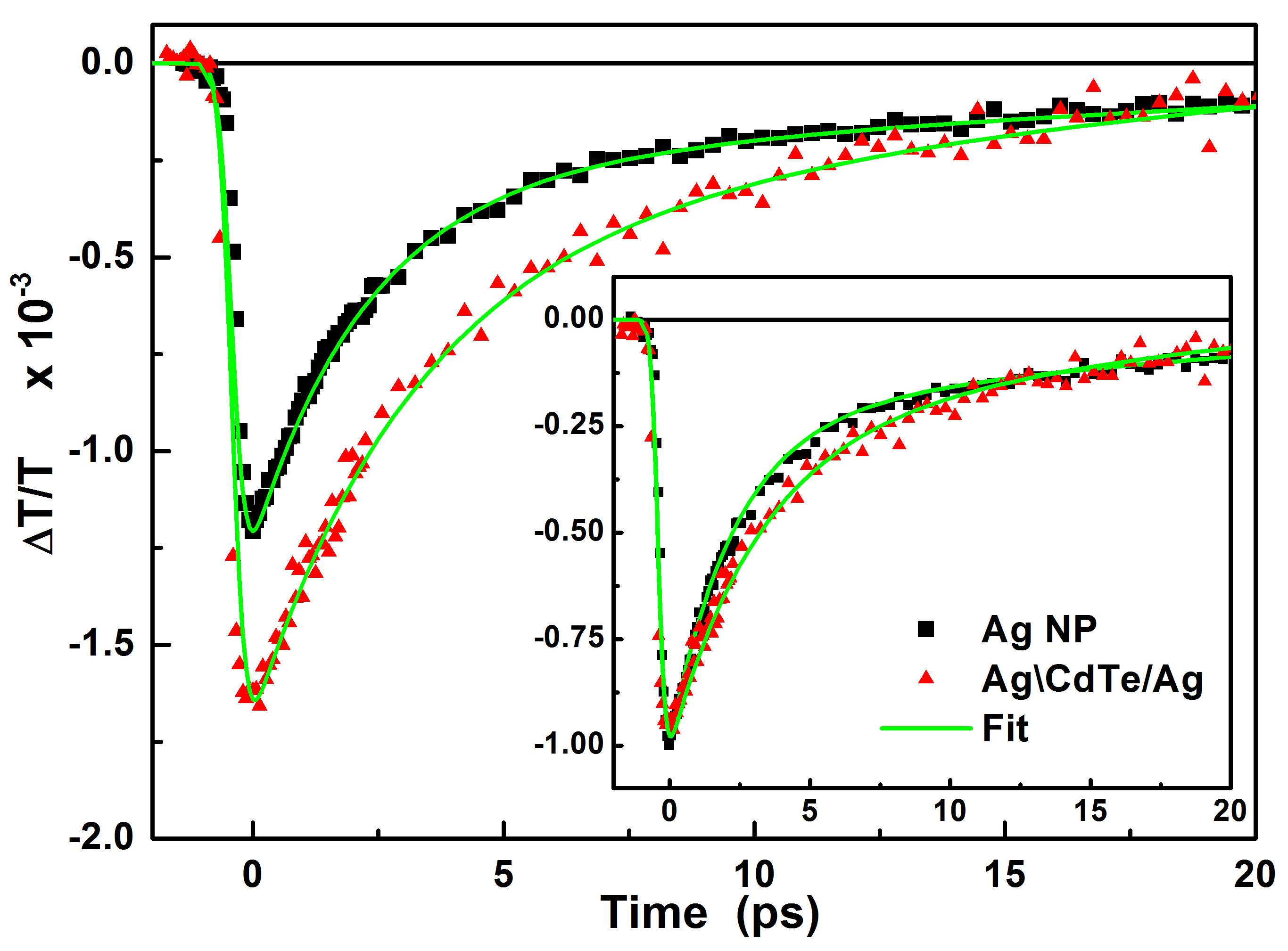}}
	\caption{Transient transmission curve of Ag NPs and Ag\textbackslash CdTe/Ag hybrid colloidal 
samples excited at 400 nm and probe at 550 nm. Inset shows the normalized transient curve of these 
samples. \label{Fig:Transient550nm}}
\end{figure}

Figure.\ref{Fig:Transient410nm} and Figure.\ref{Fig:Transient550nm} shows the transient transmission signal 
($\Delta T/T$) measured for the bare Ag NP and Ag\textbackslash CdTe/Ag hybrid when probed at 408 nm and 550 nm 
respectively. In all these measurements the pump wavelength (400 nm) and pump fluence (2.1 $\mu$Jmm$^{-2}$) at 
the sample place were kept same. For both the samples and at both probe wavelengths, the magnitude of 
$\Delta T/T$ ($|\Delta T/T|$) increases, reaching a maximum by about 400 fs. With further increase in the 
pump-probe delay, $|\Delta T/T|$ starts recovering, reaching a very low value by about 20 ps. Further, we find that
sample containing bare CdTe QD colloid of concentration similar to that used in forming the Ag\textbackslash CdTe/Ag 
hybrid colloid did not show any measurable $\Delta T/T$ signal.

The origin of ultrafast optical response of the bare metal NP colloid has been reported by several 
groups\cite{Landua-damping-guillon-2004-ultrafast, Electron_counting_Jayabalan_JPCC_2019, 
	Non-equilibrium-electron-dynamics-nobel-PRB-2000, Electron-dynamics-Chemical-Physics-2000, 
	Higher-nonlinearity-JOSAB-Jayabalan_2011}. When a ultrashort pulse excites 
a metal nanoparticle at its LSPR, the free-electrons in the particles are set to oscillate in phase 
with the applied field. Within next few femtoseconds, these electrons decay to single-particle states 
through Landu-damping\cite{Landua-damping-guillon-2004-ultrafast}. The energy distribution among electrons 
at this stage will be non-thermal. These electrons relaxes to a thermalized high temperature state mainly 
through electron-electron scattering by about few hundreds of 
femtoseconds\cite{Non-equilibrium-electron-dynamics-nobel-PRB-2000}. With increase in temperature of free-electrons, 
the real part of dielectric constant of Ag also increases which red-shifts the LSPR 
peaks\cite{Higher-nonlinearity-JOSAB-Jayabalan_2011, Non-equilibrium-electron-dynamics-nobel-PRB-2000, 
	Electron-dynamics-Chemical-Physics-2000}. This shift causes a change in the 
transmission of a metal NP colloid. The peak change in the transient transmission occur at the time when 
there is a maximum change in temperature of the free-electrons. In the present case of bare Ag NP colloid, when 
excited at 400 nm, the peak in $|\Delta T/T|$ occurs nearly after 400 fs irrespective of probe wavelength. 
Further, the red-shift of LSPR will cause an increase in transmission, when probed at 408 nm (blue-side 
of LSPR), and reduction in transmission when probed at 550 nm. Thus, the difference in probing wavelength 
with respect to LSPR causes the change in the sign of the measured $\Delta T/T$ in the case of Ag NP colloid
(Fig.\ref{Fig:Transient410nm} and Fig.\ref{Fig:Transient550nm}). 

At the end of thermalization, the temperature of free-electrons is much higher whereas the lattice still 
remains almost at room temperature. Over a time of next few picoseconds, electron-phonon interaction leads 
to thermalization of electrons and lattice to a much lower temperature. This is due to the fact that the 
specific heat capacity of lattice is much higher than that of electrons. The thermalization process of 
electrons and lattice can be quantitatively understood using two-temperature 
model\cite{Higher-nonlinearity-JOSAB-Jayabalan_2011, Non-equilibrium-electron-dynamics-nobel-PRB-2000, 
	Electron-dynamics-Chemical-Physics-2000}. However, the two-temperature model would 
not be, in principle, directly applicable for hybrid nanostructures. In the present case, because hybrid 
nanostructures are also understudy, the complete temporal evolution of $\Delta T/T$ was fitted to,
\begin{equation}
F(t) = \frac{1}{2} \left[\mathcal{E}\left( t/t_r \right)+1\right] \left[ \sum_{k} A_k e^{-t/t_k} \right]
\label{Eq:Fitting}
\end{equation}
where $\mathcal{E}(x)$ is the Gauss error function of $x$, $A_k$ and $t_k$ are the amplitude and decay 
time respectively. The best fit to the experimental data using Eq.\ref{Eq:Fitting} is also shown in 
Fig.\ref{Fig:Transient410nm} and Fig.\ref{Fig:Transient550nm}. In the case of the bare Ag NP colloid, 
the best fit value of electron-phonon thermalization time ($\tau_{ep}$) probed at 408 nm and 550 nm 
are 2.3 ps $\pm 100 fs$ and 2.4 ps $\pm 100 fs$ respectively.  

Similar to the bare Ag NP colloid, the sign of $\Delta T/T$ of Ag\textbackslash CdTe/Ag hybrid colloid 
also changes from positive to negative when probe wavelength is changed from 408 nm to 550 nm. 
This shows that the transient response of Ag\textbackslash CdTe/Ag is strongly similar to that of pure Ag NP colloid. To compare the temporal response, $\Delta T/T$ normalized to the peak change in $|\Delta T/T|$ ($|\Delta T/T|_{pk}$) 
are also shown in the inset of Fig.\ref{Fig:Transient410nm} and Fig.\ref{Fig:Transient550nm}. The best 
fit to the decay times of $\Delta T/T$ for Ag\textbackslash CdTe/Ag hybrid colloid when probed at 408 
nm and 550 nm are 2.6 ps $\pm 100 fs$ and 3.2 ps $\pm 100 fs$ respectively. Clearly, when probed at 408 nm the recovery of change in 
transmission of the Ag\textbackslash CdTe/Ag hybrid colloid is slightly slower than Ag NP colloid whereas
at 550 nm it takes much longer to recover. 

Electron-phonon relaxation takes a longer time in a single crystal Ag NP compared to that of crystal having 
twin defects because of the additional scattering of electrons at the lattice 
defects\cite{Tailoring_Properties_crystanality_Nature_2007}. In the present
case, the lattice quality of the bare Ag NPs and Ag\textbackslash CdTe/Ag hybrid colloids should be same 
because the later is derived from the former. Further, the simple linking of Ag NPs by the CdTe
QD should not change its crystal quality. Thus the increase in the relaxation time for the Ag\textbackslash 
CdTe/Ag hybrid colloid cannot be explained by an improvement in lattice quality. At low temperatures, the 
electron-phonon relaxation time also depends on its temperature itself, increasing linearly with 
it\cite{Electron_counting_Jayabalan_JPCC_2019}. Thus if the increase in the temperature of Ag in the 
Ag\textbackslash CdTe/Ag hybrid nanostructure is more, then it can also show an increased electron-phonon 
relaxation time. The comparison between the measured $|\Delta T/T|_{pk}$ for different samples will be 
able to give some information about the maximum temperature reached by the free-electrons in the Ag NPs. 

If $\Delta \varepsilon^{'}_{m}$ is the small change in real part of dielectric constant due to the
change in electron temperature, then the change in transmission at the probe wavelength ($\Delta T/T$) 
of a colloid having randomly oriented ellipsoidal particle can be written 
as\cite{Higher-nonlinearity-JOSAB-Jayabalan_2011, Electron_counting_Jayabalan_JPCC_2019},
\begin{eqnarray}
\frac{\Delta T}{T} = \frac{2 \pi p}{3 \lambda_{pr} \sqrt{\varepsilon_s}} \sum_{j =x,y,z}
Im \left[ f^2_j (\lambda_{pr}) \right] \Delta \varepsilon^{'}_m (\lambda_{pp},\lambda_{pr}).\label{Eq:DelTbyT}
\end{eqnarray}
where $p$ is the volume fraction, $\varepsilon_s$ is the dielectric constant of the surrounding medium,
$f_j$ is the local field factor for $j^{th}$ principle axis\cite{Absorption_Scattering_small_particles_borhen_1983}. Because the change 
in dielectric constant is induced by the pump pulse and is sensed at probe wavelength, $\Delta \varepsilon^{'}_m$ 
should depend on both the pump and probe wavelengths and their polarizations with respect to the orientation 
of the particle. For sufficiently small absorbed pump energies, $\Delta \varepsilon^{'}_m$ depends linearly 
on absorbed power per unit volume of the particle\cite{Higher-nonlinearity-JOSAB-Jayabalan_2011}. For small 
particles, the contribution to the extinction cross-section is mainly dominated by absorption. 
Thus, $\Delta \varepsilon^{'}_m$ is expected to be directly proportional to the extinction cross-section of 
the particle. 

\begin{figure}[h] 
	\centerline{\includegraphics[width=0.95\columnwidth]{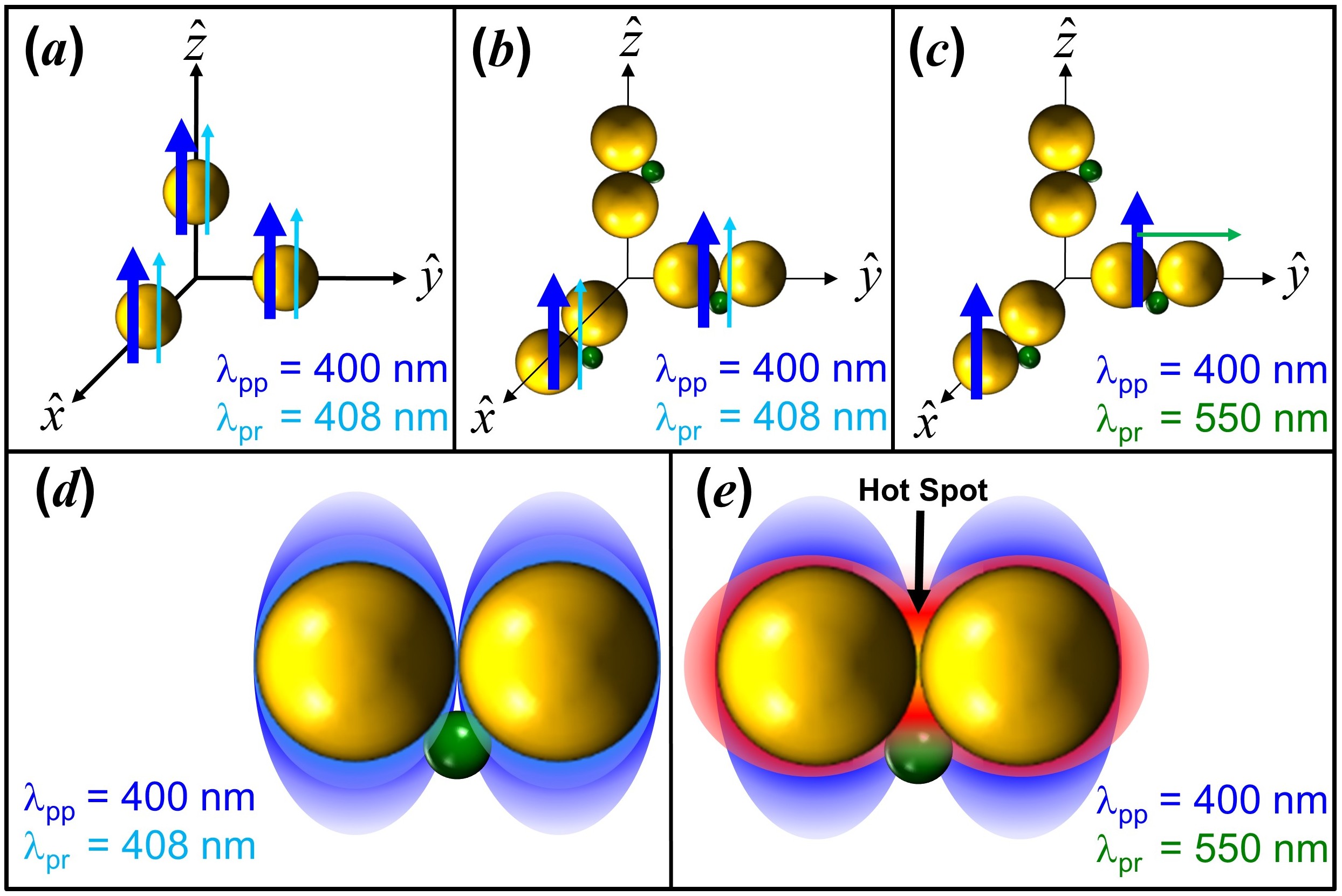}}
	\caption{Schematic of particles oriented in three different directions along with polarization of 
		pump and probe pulse. Blue thick line indicates polarization of pump pulse(400 nm) and thin line, 
		blue(408 nm) and green(550 nm), indicates polarization of probe pulse. 
		\label{Fig:OrientationsTransient}}
\end{figure}

The $|\Delta T/T|_{pk}$ of Ag NP colloid when probed at 550 nm is much lower (only 9\%) than that when probed 
at 408 nm. When pumped at 400 nm, irrespective of the polarization, all the Ag NPs absorbs the light. The change
in the temperature of free-electrons will be proportional to the $C_{ext}$ of a single Ag NP. When probed at 408 nm, 
all the Ag NPs will contributes to the $|\Delta T/T|_{pk}$ and the contribution from $Im \left[ f^2_j \right]$ 
is also strong near to the LSPR (Fig.\ref{Fig:OrientationsTransient} {\it a}). However, when probed at 550 nm which is far separated from LSPR, the contribution to $|\Delta T/T|_{pk}$ from $Im \left[ f^2_j \right]$ reduces 
considerably\cite{Higher-nonlinearity-JOSAB-Jayabalan_2011, Hot_electron_metal_PRL_2000,
	Experimental_investigation_AgNP_whitelight_PRB_2014}, reducing the measured $|\Delta T/T|_{pk}$. Because
the pump fluence is same in both of these case the decay time will also remain same irrespective of
the probe wavelength.

Consider the simple model of Ag\textbackslash CdTe/Ag particles oriented along the three Cartesian directions. 
In Fig.\ref{Fig:OrientationsTransient}, schematic of particles oriented in three different directions along 
with the polarizations of the pump and probe beams used in the experiments are shown. As discussed in the 
case of static optical response, when the particle is aligned along the transverse axis, its $C_{ext}$ will remain 
nearly same as that of two isolated Ag NPs. Thus when excited at 400 nm, close to T-LSPR, nearly 
$\sfrac{2}{3}^{rd}$ of the Ag\textbackslash CdTe/Ag hybrid particles will absorb the light and the increase 
in temperature of free-electrons will also be nearly same as that of the isolated Ag NP for the same pump 
fluence. When probed at 408 nm, the factor $Im \left[ f^2_j \right]$ will also be finite and all the 
excited Ag\textbackslash CdTe/Ag particles (which is only $\sfrac{2}{3}^{rd}$ of the total) will contribute 
to $|\Delta T/T|_{pk}$. Thus the $|\Delta T/T|_{pk}$ measured
for Ag\textbackslash CdTe/Ag hybrid colloid should be lesser by at least a factor of $\sfrac{2}{3}^{rd}$ 
compared to that of Ag NP colloid probed at 408 nm. With a similar argument it can be shown that the 
$|\Delta T/T|_{pk}$ of Ag\textbackslash CdTe/Ag hybrid colloid probed at 550 nm, close to L-LSPR,
should be lesser by at least a factor of $\sfrac{1}{3}^{rd}$ compared to that of Ag NP colloid probed 
at 408 nm (Fig.\ref{Fig:OrientationsTransient}{\it c}). The $|\Delta T/T|_{pk}$ measured for the 
Ag\textbackslash CdTe/Ag hybrid colloid is about $\sim$ 44\% and 11\% when probed at 408 nm and 550 nm 
respectively, compared to that of Ag NP colloid at 408 nm. Thus, the measured $|\Delta T/T|_{pk}$ of 
Ag\textbackslash CdTe/Ag hybrid colloid at these wavelengths were even lower than that estimated. 
Based on this observation, the increase in temperature of free-electrons in the Ag NPs of 
Ag\textbackslash CdTe/Ag hybrids should be same or lower than that of the particles in bare 
Ag NP colloid. Thus the electron-phonon relaxation time measured for Ag\textbackslash CdTe/Ag hybrid 
colloid should have remained same or reduced than that of Ag NP colloid, which is opposite to that of 
observed in the experiment.
Hence, change in the absorbed power cannot explain the increase in the relaxation time measured for
Ag\textbackslash CdTe/Ag hybrid colloid.

Prashant K. Jain {\it et al.} studied the transient change in the optical response of aggregated 
gold nanoparticles when excited at 400 nm\cite{Ultrafast_electron_relaxation_dynamics_aggregates_JPCB_2006}. 
Their result show that the electron-phonon relaxation time is smaller when probed at the longer-wavelength 
LSPR peak compared to that measured at the shorter wavelength. They have attributed this reduction to 
the increased overlap of the longer wavelength LSPR with the phonon spectrum and also to the enhanced 
interfacial electron scattering\cite{Ultrafast_electron_relaxation_dynamics_aggregates_JPCB_2006}. 
Similar reduction in electron-phonon relaxation time has also been reported for other Au and Ag 
nanoparticle aggregates\cite{feldstein1997electronic, yang2009effects}. This implies that the interaction 
between two metal nanoparticles should lead to a reduction in the relaxation time of the $\Delta T/T$, 
which is again opposite of what is observed in the present case. Thus, presence of CdTe QD in between 
the Ag nanoparticles is indeed playing a role in the dynamics observed in Ag\textbackslash CdTe/Ag colloid.

When a metal nanoparticle is attached to a semiconductor, three different processes starts
modifying the response of the combined super-structure. A direct metal-to-semiconductor interfacial charge 
transfer, which occurs through direct excitation of electron from metal NPs to empty states in 
semiconductor\cite{interfacial_charge_transfer_Science_2015, Review_plasmon_hot_electron_transfer_NPG_2017,Gao2012}. Such 
transfer is more possible in hybrid systems where semiconductor is in direct contact with 
the surface of the metal. In some hybrid cases, the metal NP absorbs the light creating hot-electrons, 
which then gets transferred to the empty levels in semiconductor. Signature of such 
hot-electron transfer from metal NP to semiconductor QD has been observed in several hybrid 
nanostructures\cite{interfacial_charge_transfer_Science_2015, Review_plasmon_hot_electron_transfer_NPG_2017,
	Efficient_Ag_AgCl_plasmon_transfer_electron_transfer_Adv_2013}. Even if there is no direct contact 
between the metal and semiconductor, hot-carrier transport still occurs via the linking
polymer/molecule\cite{Samanta-ultrafast-AgCdTe-JPCC-2016, Electron_counting_Jayabalan_JPCC_2019}. 
In addition to the above mentioned processes,
a direct energy transfer between metal and semiconductor is also possible, if the LSPR of metal nanoparticle 
overlaps with the absorption spectrum of semiconductor QD\cite{Surface_energy_transfer_JPCC_2018, 
	PL_UCPL_Ragab_2014}. 

Charge or energy transfer between the constituents of a hybrid strongly depends on the final superstructure 
formed by the metal-semiconductor nanostructures\cite{Au_TiO2_charge_separation_JACS_2013, 
	Samanta-ultrafast-AgCdTe-JPCC-2016, Electron_counting_Jayabalan_JPCC_2019}. When the Ag\textbackslash CdTe/Ag hybrid 
colloid is pumped at 400 nm, T-LSPR is excited in the Ag NPs and at the same time it can also excite carriers 
in CdTe from valance band to conduction band. A comparison of temporal response of $\Delta T/T$ of Ag NP 
colloid and Ag\textbackslash CdTe/Ag colloid when probed at 408 nm shows that the dynamics in this system 
is dominated by the plasmonic response of a Ag NP. Even if carriers are excited in CdTe QDs, when 
probed well above the band edge, CdTe QDs do not show significant change in its transient
response\cite{Samanta-ultrafast-AgCdTe-JPCC-2016}. Further, exciting at 400 nm, resonant to T-LSPR,
does not create "hot-spot" in between the Ag NPs where CdTe QD is located 
(Fig.\ref{Fig:OrientationsTransient}{\it d}). Thus probing the Ag\textbackslash CdTe/Ag colloid at 408 nm 
strongly resembles the dynamics of Ag NP colloid (Fig.\ref{Fig:Transient410nm}). On the other hand, when 
probed at 550 nm, the probe pulse can sense the presence of CdTe QD because probe field gets enhanced at the 
location of CdTe QD. Further, CdTe QD can also show a strong transient change in transmission when probed 
at 550 nm. Thus, although a significant contribution of metallic response is expected near to the L-LSPR, the 
Ag\textbackslash CdTe/Ag hybrid colloid, when probed at 550 nm show a longer response time than that of 
bare Ag NP colloid. 

\section{Conclusion}
When Ag NP colloid is mixed with TGA capped CdTe QD colloid such that in the mixed colloid 
the number ratio between the Ag NPs and CdTe QDs is 2:1, stable hybrid nanostructure with two Ag 
NPs and a CdTe could be formed.  Optical spectra during growth and transient optical response 
suggests that the final structure has two touching Ag nanoparticle with CdTe QD stuck in between 
them. The static optical response of the final hybrid colloid could be 
explained well by the electromagnetic interaction between these particles present in the structure. 
Assuming that the extinction cross-section of each Ag NP in the Ag\textbackslash CdTe/Ag hybrid 
structure to be same as that of an isolated Ag NP, the optical response of Ag\textbackslash CdTe/Ag 
colloid could be explained by their random orientations in the host.
When excited and probed near the T-LSPR, the ultrafast optical 
response of Ag\textbackslash CdTe/Ag hybrid nanostructure resembles that of a bare Ag NP. On the other hand while 
probing at the L-LSPR show a delayed recovery of transient transmission. This is attributed 
to the sensing of presence of CdTe QD in the hot-spot regime by the probe field. Placing a 
semiconductor QD at or close to the hot-spot is a major step towards using the enhanced local 
field in various applications such as plasmonic sensing, optoelectronics, energy harvesting, 
nanolithography, and optical nano-antennas. Further, these studies are also important for the 
advancement in the area of colloidal self-assembly, with impacts on the dynamic properties of 
hot spots for specific needs\cite{Hot_spot_SERS_PCCP_2015, Hot_Spot_ultrafast_plasmon_Nature_2017}.

\section*{Acknowledgments}
The authors are thankful to Tarun Kumar Sharma, Rama Chari and Salahuddin Khan for fruitful discussions and 
suggestions. The authors are also thankful to Arvind Kumar Srivastava for the support in obtaining
the TEM images. The authors, Sabina Gurung and Durga Prasad Khatua are thankful to RRCAT, Indore, under 
HBNI programme, Mumbai, for the financial support.

\section*{References}
\bibliographystyle{unsrt}
\bibliography{Article-AgCdTeAg.bib}


\end{document}